\documentclass[aps,twocolumn ,nofootinbib,superscriptaddress,geometry,preprintnumbers]{revtex4-1}

\usepackage{amsmath,amssymb} 
\usepackage{accents}
\usepackage{nccmath}
\usepackage{color}
\usepackage[bookmarks=true,hyperfigures=true]{hyperref}
\usepackage{graphicx}
\usepackage[export]{adjustbox}
\usepackage{tensor}
\usepackage{empheq}
\usepackage{nicefrac}
\usepackage{shuffle}
\usepackage{slashed}
\usepackage{subfigure}

\usepackage{feynmp-auto}
\usepackage{feynmp} 
\allowdisplaybreaks[3]

\usepackage[font=small,labelfont=bf,width=0.85\textwidth]{caption}

\let\oldbfseries=\bfseries
\let\oldmdseries=\mdseries
\let\oldnormalfont=\normalfont
\renewcommand{\bfseries}{\oldbfseries\boldmath}
\renewcommand{\mdseries}{\oldmdseries\unboldmath}
\renewcommand{\normalfont}{\oldnormalfont\unboldmath}

\makeatletter
\newlength{\apb@width}
\newcommand{\autoparbox}[2][c]{\settowidth{\apb@width}{#2}\parbox[#1]{\apb@width}{#2}}

\makeatother

\newcommand{\nn}{\nonumber}

\newcommand{\remark}[2][.]{{\color{red}\renewcommand{\bfdefault}{b}\rmfamily\if.#1\else\textbf{#1:} \fi#2}}

\newcommand{\ee}{\end{equation}}
\newcommand{\beq}{\begin{equation}}
\newcommand{\eeq}{\end{equation}}

\newcommand{\bma}{\begin{pmatrix}}
\newcommand{\ema}{\end{pmatrix}}
\newcommand{\ba}{\begin{eqnarray}}
\newcommand{\ea}{\end{eqnarray}}
 
\ifx\genfrac\sdflkaj\else\fi

\newcommand{\cO}{\mathcal{O}}

\def\l<{\langle}\def\r>{\rangle}

\makeatletter
\newcommand{\namedref}[2]{\hyperref[#2]{#1~\ref*{#2}}}
\newcommand{\secref}{\@ifstar{\namedref{Section}}{\namedref{sec.}}}
\newcommand{\subsecref}{\@ifstar{\namedref{Subsection}}{\namedref{subsec.}}}
\newcommand{\appref}{\@ifstar{\namedref{Appendix}}{\namedref{app.}}}
\newcommand{\tabref}{\@ifstar{\namedref{Table}}{\namedref{tab.}}}
\newcommand{\figref}{\@ifstar{\namedref{Figure}}{\namedref{fig.}}}
\makeatother

\def\[{\begin{equation}}
\def\]{\end{equation}}
\def\<{\begin{eqnarray}}
\def\>{\end{eqnarray}}

\newcommand{\eqn}[1]{(\ref{#1})}

\def\bea{\begin{align}}
\def\eea{\end{align}}
\def\be{\begin{equation}}
\def\ee{\end{equation}}

\begin{document}
\newtheorem{theorem}{Theorem}
\newtheorem{theorem1}{Theorem}
\thispagestyle{empty}


\title{The Double Copy of Massive Scalar-QCD}
\author{Jan Plefka}
\affiliation{Institut f\"ur Physik und IRIS Adlershof, Humboldt-Universit\"at zu Berlin, 
Zum Gro{\ss}en Windkanal 6, 12489 Berlin, Germany}
\author{Canxin Shi}
\affiliation{Institut f\"ur Physik und IRIS Adlershof, Humboldt-Universit\"at zu Berlin, 
Zum Gro{\ss}en Windkanal 6, 12489 Berlin, Germany}
\author{Tianheng Wang}
\affiliation{Institut f\"ur Physik und IRIS Adlershof, Humboldt-Universit\"at zu Berlin, 
Zum Gro{\ss}en Windkanal 6, 12489 Berlin, Germany}
\email{jan.plefka@physik.hu-berlin.de, canxin.shi@physik.hu-berlin.de, tianheng.wang@physik.hu-berlin.de}
\preprint{HU-EP-19/32, SAGEX-19-28-E}

\begin{abstract}
We construct the gravitational theory emerging from the double-copy of
massive scalar quantum chromodynamics in general dimensions. 
The resulting two-form-dilaton-gravity theory couples to flavored massive scalars gravitationally and via the dilaton. It displays scalar self-interaction 
terms of arbitrary even order in the fields but quadratic in derivatives. We
 work out the emerging Lagrangian explicitly up to the sixth order in scalar fields
 and propose an all order form.
\end{abstract}

\maketitle

\section{Introduction}
%
%
%
%

A most remarkable property of 
gravitational scattering amplitudes on Minkowski space-time backgrounds is a 
factorization of these highly involved structures into simpler 
building blocks arising from two gauge theoretical scattering amplitudes known as the Bern-Carrasco-Johansson (BCJ) double-copy relation \cite{Bern:2008qj,Bern:2010ue} \footnote{See \cite{Bern:2019prr} for a recent comprehensive review.}. The BCJ double-copy acts inherently perturbatively and reflects a fascinating duality between the gauge theoretic 
color and kinematical degrees of freedom of scattering amplitudes. 
It provides a mechanism
to transform the local gauge symmetries of the double-copied Yang-Mills theory to the
diffeomorphism symmetry of the resulting gravitational theory. 
This procedure hinges upon finding a representation of the gauge theory amplitudes 
built from trivalent graphs with color and kinematic numerator factors fulfilling 
Jacobi-like identities. It was originally observed for pure gauge theories 
(with and without supersymmetry) which double copied to pure (super)-gravities 
and has proven to be
an efficient tool also beyond tree-level (where proofs of the double-copy exist \cite{Bern:2010yg,Tye:2010dd}) 
to generate amplitude integrands at high-loop orders in the maximal supersymmetric case \cite{Bern:2017ucb,Bern:2018jmv}. 
By now the double-copy has been observed in a large set of gravitational theories involving massless fields, including pure gravity \cite{Johansson:2014zca}, pure and matter coupled supergravities
\cite{Johansson:2014zca,Johansson:2017bfl,Carrasco:2012ca} as well as Einstein-Yang-Mills theories \cite{Chiodaroli:2014xia,Chiodaroli:2015wal,Chiodaroli:2017ehv,Faller:2018vdz}. In \cite{Johansson:2015oia,Johansson:2019dnu},
 the double
copy of QCD with $N_{f}$ \emph{massive} spin $1/2$ particles in the
fundamental representation was studied following , which double copies to two-form-dilaton-gravity (a.k.a.~$\mathcal{N}=0$ supergravity) coupled to massive scalars and vectors, see also \cite{Bautista:2019evw} for related work. 
This field content of the double copied theory follows from the tensor product of the on-shell
degrees of freedom
\be
[\,A_{\mu} \oplus (\,N_{f}\times \Psi \,)\, ]^{\otimes 2}
= h_{\mu\nu} \oplus B_{\mu\nu} \oplus \phi \oplus (N_{f}\times[ \phi \oplus V^{\mu}]) \, .
\ee
The resulting gravitational Lagrangian is not obvious at all in the
matter sector and contains contact terms of pure scalar and vector-scalar type
to all orders. The explicit form up to
quartic order was worked out in detail in \cite{Johansson:2019dnu}. 

In this work we extend these results by going down in spin and work out
the details of the double copy of \emph{massive scalar} QCD. In this setup the resulting 
field content
is two-form-dilaton-gravity  coupled to $N_{f}$ massive scalars, i.e.
\be
[\,A_{\mu} \oplus (\, N_{f}\times \varphi \, )\, ]^{\otimes 2}
= h_{\mu\nu} \oplus B_{\mu\nu} \oplus \phi \oplus ( N_{f}\times \varphi  )\, .
\ee
We have constructed the emerging gravitational Lagrangian up to \emph{sixth order} in the
scalar contact terms and moreover propose a general structure. We also show that there are no 
two-form-scalar couplings, i.e. axion-scalar couplings in 4d, as is commonly expected. Partial results along these
lines up to quartic order in scalar couplings appeared recently in \cite{Bautista:2019evw} as this work was close to completion.

Next to clarifying the general structure of the double-copy construction  
one further  motivation for this work lies the application to the
\emph{classical} double copy. This is particularly 
relevant in applications to the two-body problem
in general relativity -- being of central importance for gravitational wave physics.
Here the goal is to use the double-copy construction to innovate state-of-the-art
techniques for the computation of effective two-body potentials in the post-Newtonian and
post-Minkowski expansions which is an emerging interdisciplinary field \cite{Goldberger:2016iau,Luna:2016hge,Goldberger:2017vcg, Goldberger:2017ogt,Goldberger:2017frp,Chester:2017vcz,Shen:2018ebu,Luna:2017dtq,Plefka:2018dpa,Guevara:2018wpp,Chung:2018kqs,Bern:2019nnu,Bern:2019crd,Kosower:2018adc,Maybee:2019jus,Arkani-Hamed:2019ymq,Chung:2019duq,Guevara:2019fsj,Cristofoli:2019neg,Kalin:2019rwq,Bjerrum-Bohr:2019kec,Brandhuber:2019qpg,Bjerrum-Bohr:2019nws}. This connection 
comes about from an effective field theory perspective where spinless black-holes may be modeled by massive scalar particles in the classical and large mass 
limit (for introductory reviews see \cite{Goldberger:2007hy,Porto:2016pyg,Levi:2018nxp}). 
While the double copy procedure could be
efficiently applied up to the 1PN and 2PM level \cite{Goldberger:2016iau,Goldberger:2017vcg, Goldberger:2017ogt,Goldberger:2017frp,Chester:2017vcz,Shen:2018ebu,Plefka:2018dpa} a recent double-copy construction at 
the 2PN and 3PM level of the present authors and Steinhoff revealed a disagreement with the 
known 2PN result in dilaton-gravity \cite{Plefka:2019hmz}.

Let us briefly review the double-copy procedure at tree-level.
A general $n$-point amplitude in the gauge theory, here scalar QCD, takes the generic
form
\be
\label{dc}
A_{n}= g^{m-2} \sum_{j\in\Gamma} \frac{1}{S_{j}}\frac{c_{j}\, n_{j}}{\prod_{\alpha_{j}}D_{\alpha_{j}}} \ee
where $\Gamma$ is the set of relevant trivalent graphs
 and $g$ the coupling. The $D_{\alpha}$ denote the inverse propagators and $S_{j}$ is the
 symmetry factor of the graph.
Note that via the trivial identity $1=D_{\alpha_{j}}/D_{\alpha_{j}}$ \emph{any} graph
in a Feynman diagrammatic expansion can be made trivalent formally.
The central criterion is that in the above representation of the amplitude the numerators of color $c_{i}$ and kinematics $n_{i}$ obey identical algebraic relations
\be
c_{s}+c_{t}=c_{u}  \qquad n_{s}+n_{t}=n_{u} 
\ee
emerging from the Jacobi identity for the $c_{i}$. The double-copied amplitude
 is then obtained by simply replacing $c_{i}\to n_{i}$ in \eqn{dc}, a process known 
 as `squaring'.

\section{Scalar-QCD amplitudes}
The scalar-QCD Lagrangian with the scalar fields in the fundamental representation reads
\begin{align}\label{eq:LsQCD}
	\mathcal{L}_{\text{sQCD}} = & -\frac{1}{4} F^a_{\mu\nu}F^{\mu\nu,a}  - \frac{1}{2} (\partial^\mu A_\mu^a)^2  \nonumber\\
	&+ \sum_\alpha \Big( (D_\mu \varphi_{\alpha,i})^\dagger D^\mu \varphi_{\alpha,i} - m_\alpha^2 \varphi_{\alpha,i}^\dagger \varphi_{\alpha,i} \Big),
\end{align}
where $\alpha=1,2,\cdots,N_f$ labels the flavor of the massive scalar fields and $i$ denotes the color index. 
We note that there are no contact interactions of massive scalar fields in this Lagrangian.
The conventions and the Feynman rules following from \eqref{eq:LsQCD} are given in Appendix \ref{appendix:FeynRules}.

From the Feynman rules, the 3-point amplitude of two massive scalars and one gluon can be read off immediately,
\begin{align}\label{eq:A2scalar1gluon}
 A(\underline{1}_{\alpha,j}, \overline{2}_{\alpha,i}, 3^a) = i {g \over \sqrt{2}} T^{a}_{ij} \epsilon_3 \cdot (p_1 - p_2),
\end{align}
where the massive scalar particle and antiparticle are indicated by the under- and over-lines respectively, $p_i$ denotes the external momentum of particle $i$ and all external momenta are incoming. 

The scattering amplitude of two gluons and two massive scalars reads
\begin{align}\label{eq:A2scalar2gluon}
	&A(\underline{1}_{\alpha,j}, \overline{2}_{\alpha,i}, 3^a, 4^b) \nonumber\\
	= &  \frac{(T^{a}T^b)_{ij} n_u}{s_{14} -m^2_\alpha} 
	+ \frac{(T^{b}T^a)_{ij} n_t}{s_{13} -m^2_\alpha} 
	+ \frac{f^{abc}T^c_{ij} n_s}{s_{34}}.
\end{align}
Note that we use $s_{i..j} \equiv (p_i + ... +p_j)^2$ throughout this paper. The kinematic numerators are given by
\begin{align}
	n_u =& ig^2 \left[ 2(\epsilon_4 \cdot p_1)(\epsilon_3 \cdot p_2) + (\epsilon_3 \cdot \epsilon_4)(p_1 \cdot p_4) \right]\\
	n_t =& ig^2 \left[ 2(\epsilon_3 \cdot p_1)(\epsilon_4 \cdot p_2) + (\epsilon_3 \cdot \epsilon_4)(p_1 \cdot p_3) \right]\\
	n_s =& ig^2 \left[ 2(\epsilon_4 \cdot p_1)(\epsilon_3 \cdot p_2) - 2(\epsilon_3 \cdot p_1)(\epsilon_4 \cdot p_2)  \right. \nonumber\\
& \qquad \left.+ (\epsilon_3 \cdot \epsilon_4) p_1 \cdot (p_4 - p_3) \right].
\end{align}
We notice that the color and the kinematic numerators above automatically satisfy the relations below
\begin{align}
	(T^{a}T^b)_{ij} &- (T^{b}T^a)_{ij} = f^{abc}T^c_{ij},\\
	n_u &- n_t = n_s.
\end{align}
Hence, the kinematic numerators are in the BCJ-respecting representation and are ready to be squared in the double copy construction. 

The 4-scalar amplitudes can also be straightforwardly obtained. When the two pairs of scalars are of the same flavor (mass), the amplitude reads
\begin{align} \label{eq:4scalars}
	&A({\color{red} \underline{1}_{\alpha,j}},{\color{red} \overline{2}_{\alpha,i}},{\color{red} \underline{3}_{\alpha,l} },{\color{red} \overline{4}_{\alpha,k}} )  \\
	= & T^a_{ij}T^a_{lk} \frac{i g^2 [ p_3 \cdot (p_1 - p_2)]}{s_{34}} 
	+ T^a_{ik}T^a_{lj} \frac{i g^2 [ p_3 \cdot (p_1 - p_4)]}{s_{14}}. \nonumber
\end{align}
When the two scalar pairs are of different flavors (masses), the amplitude reads
\begin{align} \label{eq:2scalars2scalars}
	A({\color{red} \underline{1}_{\alpha,j} },{\color{red} \overline{2}_{\alpha,i}} , {\color{blue} \underline{3}_{\beta,l} },{\color{blue} \overline{4}_{\beta,k}})  =  T^a_{ij}T^a_{lk} \frac{i g^2 [ p_3 \cdot (p_1 - p_2)]}{s_{34}}.
\end{align}
In these 4-scalar amplitudes, there are not enough color structures to form Jacobi identities. 

\begin{figure}[tbp!]
	\begin{subfigure}{}
		\includegraphics[scale=1]{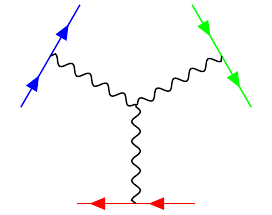} 
	\end{subfigure}
	\begin{subfigure}{}
		\includegraphics[scale=1]{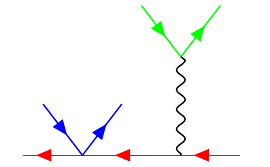} \\ 
	\end{subfigure}
	\begin{subfigure}{}
		\includegraphics[scale=1]{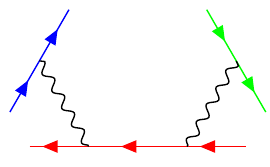} 
	\end{subfigure}
	\begin{subfigure}{}
		\includegraphics[scale=1]{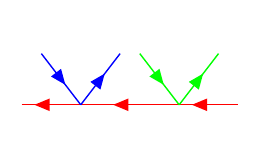} \\ \ 
	\end{subfigure}
	\begin{subfigure}{}
		\includegraphics[scale=1]{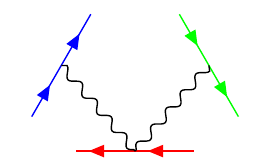} 
	\end{subfigure}
	\begin{subfigure}{}
		\includegraphics[scale=1]{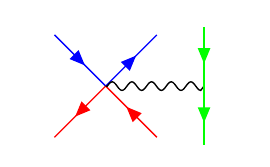} 
	\end{subfigure}
	
	\caption{Topologies of Feynman diagrams of the 6-scalar \\amplitude. Curly lines denotes possible massless propagators.}
	\label{fig:image2}
\end{figure}

The simplest 6-scalar amplitude is the one with three scalar pairs of distinct flavors (masses). 
The three topologies of the diagrams that contribute to this amplitude are given in the left column of figure \ref{fig:image2} and the amplitude reads
\begin{align} \label{eq:6scalars}
 & A( {\color{red} \underline{1}_{\alpha,j}} ,{\color{red} \overline{2}_{\alpha,i}} , {\color{blue} \underline{3}_{\beta,l} },{\color{blue} \overline{4}_{\beta,k}},{\color{green} \underline{5}_{\kappa,n}}, {\color{green} \overline{6}_{\kappa,m}}) \nonumber\\
=& {c_0 n_0 \over s_{12} s_{34} s_{56} } + \bigg[ {c_1 n_1 \over \left(s_{134} - m_\alpha^2\right) s_{34} s_{56}  }\nonumber\\
& + {c_2 n_2 \over \left( s_{156} - m_\alpha^2 \right) s_{34} s_{56} } + \big(\text{cyclic} \big) \bigg].
\end{align}
where the color factors read
\begin{align}
c_0 &= T^a_{ij} T^b_{kl} T^c_{mn} f^{abc}, \nonumber \\
c_1 &= T^b_{ih} T^c_{hj} T^b_{mn} T^c_{kl}, \\
c_2 &= T^c_{ih} T^b_{hj} T^b_{mn} T^c_{kl}.\nonumber
\end{align}
The other numerators are gained from cyclic rotations of the labels $(1,2)\rightarrow (3,4) \rightarrow (5,6)$ in $n_1$ and $n_2$.
The explicit expressions for the kinematic numerators are given in Appendix \ref{appendix:FeynRules}. We note that the color factors and the kinematic numerators satisfy the same relation below
\begin{align}
c_1 - c_2 = c_0\,,\qquad n_1-n_2 = n_0\,.
\end{align}
For the cases where the flavors are not all distinguished, more diagrams and the symmetry factors associated with the diagrams need to be taken care of. In the end, the BCJ-respecting numerators are likewise constructed. 

\section{Double copy}\label{sec:DC}

It is well-known that the double copy of YM field gives a graviton, a two-form field (B-field), and a dilaton. Their associated polarization tensors are identified with the tensor products of the gluon polarization vectors in the following ways, 
\begin{equation}
\begin{split}
	\text{graviton} \quad &: \quad (\epsilon^h)^{ij}_{\mu\nu}= \epsilon^{((i}_\mu \epsilon^{j))}_\nu,\\
	\text{B-field} \quad &: \quad (\epsilon^B)^{ij}_{\mu\nu}= \epsilon^{[i}_\mu \epsilon^{j]}_\nu,\\
	\text{dilaton} \quad &: \quad (\epsilon^\phi)_{\mu\nu}= \frac{\epsilon^{i}_\mu \epsilon^{j}_\nu \delta_{ij}}{\sqrt{D-2}},
\end{split}
\end{equation}
where the double parenthesis denotes taking the symmetric-traceless part of the tensor product. Note that the superscripts $i,j$ are not related to the color group as in \eqref{eq:LsQCD}, but are the little group indices. The dilaton identification can be further simplified as
\begin{align}
	(\epsilon^\phi)_{\mu\nu} (p,q) = \frac{1}{\sqrt{D-2}} \left( -\eta_{\mu\nu} + \frac{p_\mu q_\nu + p_\nu q_\mu}{p \cdot q} \right),
\end{align}
where $p$ is the momentum associated to the external particle, and $q$ is an arbitrary reference null-vector. 

With this double-copy prescription, we can now get the gravitational amplitudes from the corresponding YM ones. We focus on the scattering processes that involve at least one pair of massive scalars. We present all the relevant double copy results up to 4-point as well as the 6-scalar amplitude in this chapter. The simplest case on the YM side is the 3-point amplitude with two scalars and one gluon. Its double copy gives two different amplitudes,
\begin{gather}
	\mathcal{M}(\underline{1},\overline{2},3_h )= i\kappa (\epsilon_3^h)_{\mu\nu} p_1^\mu p_2^\nu \label{eq:A2scalar1graviton},\\
	\mathcal{M}(\underline{1},\overline{2},3_\phi) = \frac{i \kappa m^2}{\sqrt{D-2}} \label{eq:A2scalar1dilaton}.
\end{gather} 
Note that the amplitude vanishes when the massless particle is the B-field due to the anti-symmetrization. This applies to any amplitude that involves an odd number of B-fields, so we will only write down the non-vanishing ones. 
Similarly, the 4-point amplitude of two massive scalars and two gluon \eqref{eq:A2scalar2gluon} gives the following gravitational amplitudes after the double copy,  
\begin{align} \label{eq:A2massive2graviton}
	\mathcal{M}&(\underline{1},\overline{2},3_{h},4_{h}) = i \kappa ^2 (\epsilon^h_3)_{\mu\nu} (\epsilon^h_4)_{\rho\sigma} \bigg[\frac{1}{s_{34}} \bigg( s_{13} p_1^\rho p_2^\mu \eta^{\nu\sigma}  \nonumber\\
	& + s_{14} p_1^\mu p_2^\rho \eta^{\nu\sigma}\!\! - p_1^\mu p_1^\nu p_2^\rho p_2^\sigma - p_1^\rho p_1^\sigma p_2^\mu p_2^\nu + 2 p_1^\mu p_1^\rho p_2^\nu p_2^\sigma  \nonumber\\
	&+ \frac{1}{4}{ {s_{13}} s_{14} \eta^{\mu\rho} \eta^{\nu\sigma} }\bigg) -\frac{ p_1^\mu p_1^\nu p_2^\rho p_2^\sigma }{ s_{13}} -\frac{ p_1^\rho p_1^\sigma p_2^\mu p_2^\nu }{ s_{14}}\bigg],
\end{align}
\begin{align} \label{eq:A_2massive1dilaton1graviton}
	\mathcal{M}( \underline 1, \overline 2, 3_\phi,4_h) =& \frac{i \kappa^2 m^2  (\epsilon_4^h)_{\mu\nu} }{\sqrt{D-2}} \bigg( \frac{p_3^\mu p_3^\nu}{s_{34}} \\
	&\qquad + \frac{p_1^\mu p_1^\nu}{s_{14}-m^2} + \frac{p_2^\mu p_2^\nu}{s_{13}-m^2} \bigg),\nonumber \\
	 \label{eq:A2scalar2dilaton}
	\mathcal{M}( \underline 1, \overline 2,3_\phi,4_{\phi}) =&\frac{i \kappa^2 (p_1 \cdot p_3) (p_1 \cdot p_4)} { s_{34}}\\
	+\frac{-i \kappa^2 }{D-2} \bigg(& \frac{m^4}{s_{13}-m^2} + \frac{m^4}{s_{14}-m^2} +m^2 \bigg), \nonumber\\
	\label{eq:2massive2Bfield}
	\mathcal{M}( \underline 1, \overline 2, 3_B,4_B) =& \frac {i\kappa^2} {s_{34}}  (\epsilon^B_3)_{\mu\nu} (\epsilon_4^B)_{\rho \sigma} \Big(2 p_1^\mu p_2^\nu p_1^\rho p_2^\sigma  \\
	- 2 (p_1 \cdot p_3)& p_1^\rho p_2^\mu \eta^{\nu\sigma} - 2 (p_1 \cdot p_4) p_1^\mu p_2^\rho \eta^{\nu\sigma}  \nonumber\\
	&\quad - (p_1 \cdot p_3) (p_1 \cdot p_4) \eta^{\mu\rho} \eta^{\nu\sigma} \Big). \nonumber
\end{align}
The sQCD 4-scalar amplitudes, with the two pairs of particles being of identical and distinct flavor, \eqref{eq:4scalars} and \eqref{eq:2scalars2scalars} are double-copied to
\begin{align}
	\mathcal{M}( {\color{red} \underline 1_\alpha}, {\color{red} \overline 2_\alpha}, { \color{red} \underline 3_\alpha}, {\color{red} \overline 4_\alpha} ) =& -i \kappa^2  \bigg(  \frac{s_{23} + s_{34}}{16 } \\
	 - &\frac{p_1 \cdot p_3 \ p_2 \cdot p_3}{s_{34}} - \frac{p_1 \cdot p_3 \ p_3 \cdot p_4}{s_{23}} \bigg), \nonumber \\
	\mathcal{M}( {\color{red} \underline 1_\alpha}, {\color{red} \overline 2_\alpha}, { \color{blue} \underline 3_\beta}, {\color{blue} \overline 4_\beta} ) =& -i \kappa^2  \bigg( \frac{s_{34}}{16}  -\frac{p_1 \cdot p_3 \ p_2 \cdot p_3}{s_{34}} \bigg). \label{eq:M2massive2massive}
\end{align} 
The 6-scalar amplitude is also directly gained from \eqref{eq:6scalars}, though the result is too lengthy to fit in this article. 

These amplitudes will be exploited to extract the higher-order terms of the Lagrangian.

\section{Two-form-dilaton-gravity with massive scalars}\label{sec:gravity}
The Lagrangian of the two-form-dilaton-gravity reads
\begin{align} \label{eq:L_ADG}
	\mathcal{L}_{\text{adg}} = \sqrt{-g} \Big[ & -\frac{2}{\kappa^2} R + \frac{D-2}{2} g^{\mu\nu} \partial_{\mu} \phi \partial_\nu \phi \nonumber\\
	& + \frac{1}{6} e^{- 2 \kappa \phi} H_{\lambda\mu\nu} H^{\lambda\mu\nu} \Big].
\end{align}
The interactions involving massive scalars may now be extracted from the amplitudes computed above. In fact, we found that the minimal coupling of massive scalars and gravity reproduces the 2-massive-1-graviton amplitude \eqref{eq:A2scalar1graviton} and the 2-massive-2-graviton amplitude \eqref{eq:A2massive2graviton}. It is straightforward to extend to higher points, 
\begin{align} \label{eq:L_GM}
	\mathcal{L}_{\text{matter}} =& \sqrt{-g} \sum_{\alpha} \left( g^{\mu\nu} \partial_{\mu} \varphi_\alpha^\dagger \partial_\nu \varphi_\alpha - m_\alpha^2 \varphi_\alpha^\dagger \varphi_\alpha \right).
\end{align}
Adopting de Donder gauge, we write down the relevant Feynman rules, see Appendix \ref{appendix:FeynRules}. From the 4-point order, we can get the interaction terms up to quadratic order in $\kappa$. The 6-scalar amplitude gives us the 6-scalar contact term. In the end, we find the following Lagrangian from the double copy 
\begin{widetext}
\begin{equation}
\begin{split} \label{eq:L_pertubative}
	\mathcal{L}_{\text{DC}} =& \mathcal{L}_{\text{adg}} + \sqrt{-g} \bigg[ g^{\mu\nu} \partial_{\mu} \varphi_\alpha^\dagger \partial_\nu \varphi_\alpha - m_\alpha^2 e^{-\kappa \phi} \varphi_\alpha^\dagger \varphi_\alpha  
	+ \left ( \frac{\kappa^2}{32} \varphi_\alpha^\dagger \varphi_\alpha  - \frac{\kappa^4}{512} (\varphi_\alpha^\dagger \varphi_\alpha)^2  \right) D_\mu D^\mu (\varphi_\beta^\dagger \varphi_\beta)\bigg],
\end{split}
\end{equation}
\end{widetext}
where $D_\mu$ is the covariant derivative and flavor indices that appear twice in a term are summed over implicitly. It is also known that the dilaton appears as an exponent when coupled to other fields. 
Let us now explain the extraction of each contact term in \eqref{eq:L_pertubative}. We first make an ansatz on how the dilaton couples to massive scalars,
\begin{align}\label{eq:L_ansatz}
	\mathcal{L}_{\phi \varphi^\dagger \varphi} = \sqrt{-g} \left( e^{\lambda \kappa \phi} g^{\mu\nu} \partial_{\mu} \varphi_\alpha^\dagger \partial_\nu \varphi_\alpha -m_\alpha^2 e^{\zeta \kappa \phi} \varphi_\alpha^\dagger \varphi_\alpha \right),
\end{align}
whose leading term in $\kappa$ gives the canonical kinetic and mass terms. We can compute the 3-point amplitude of two matter fields and a dilaton,
\begin{align}
	 \mathcal{M}(\underline{1},\overline{2},3_\phi) = \frac{i(\lambda - \zeta) \kappa m^2}{\sqrt{D-2}}.
\end{align}
Matching to the double copy result  \eqref{eq:A2scalar1dilaton}, we have $\lambda-\zeta = 1$. The 2-massive-2-dilaton amplitude from double copy \eqref{eq:A2scalar2dilaton} subtracted by all Feynman diagrams that contain only cubic vertices yields the diagram that comes from the contact term, 
\begin{align} \label{eq:2massive2dilaton}
	\includegraphics[valign=c,scale=0.6]{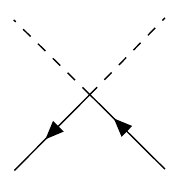} = \frac{-i \kappa^2}{2} \left( m^2( 1+ 2\lambda\zeta-2 \lambda^2) + {\lambda^2 \over 4} s_{34} \right).
\end{align}
We can also calculate this diagram directly from Feynman rules of the 4-point contact term, 
\begin{align} \label{eq:2massive2dialtonAnsatz}
	\frac{-i \kappa^2}{2} \left( m^2( \zeta^2 - \lambda^2) + {\lambda^2 \over 2} s_{34} \right).
\end{align}
Comparing the above formulae, \eqref{eq:2massive2dilaton} and \eqref{eq:2massive2dialtonAnsatz}, we have two equations of $\lambda$ and $\zeta$. Together with $\lambda-\zeta = 1$, we find the solution,
\begin{align}
	\lambda = 0, \qquad \zeta = -1.
\end{align}
We also verified that \eqref{eq:L_ansatz} gives the correct 4-point amplitude of 2 massive scalars, 1 dilaton and 1 graviton, of \eqref{eq:A_2massive1dilaton1graviton}. As for the 2 massive scalar and 2 B-field scattering, we compute the amplitude from the perturbative Lagrangian \eqref{eq:L_pertubative}. It coincides with \eqref{eq:2massive2Bfield}, so there is no direct coupling of the B-field and massive scalars at least at 3- and 4-point levels.

We now proceed to consider the interactions among the massive scalars. These will be presented in more details since it is the main result of this article. The amplitude of 2 pairs of massive scalars of different flavors contains contributions from the following diagrams,
\begin{gather}
	\includegraphics[valign=c, scale=0.6]{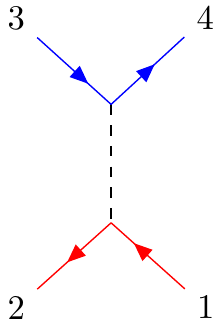} = \frac{-i m_1^2 m_2^2 \kappa^2}{(D-2) s_{34}},\\
	\includegraphics[valign=c, scale=0.6]{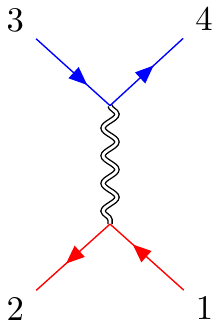} = i\kappa^2  \bigg( \frac{m_1^2 m_2^2 }{(D-2) s_{34}} + \frac{p_1 \cdot p_3 \ p_2 \cdot p_3}{s_{34}} \bigg).
\end{gather}
In order to match the double copy amplitude \eqref{eq:M2massive2massive}, a 4-scalar contact term is required,
\begin{align}
	\includegraphics[valign=c, scale=0.6]{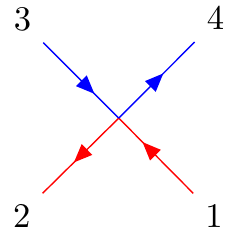} = -i \kappa^2 \frac{ s_{34} }{16}.
\end{align}
A similar calculation applies to the 4-scalar amplitude where all massive scalars are of the same flavor. We only need to add the contribution by exchanging $p_1$ and $p_3$. The contact term will be
\begin{align}
	\includegraphics[valign=c, scale=0.6]{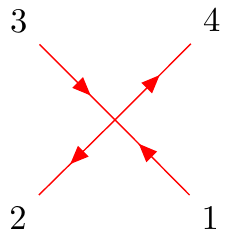} = -i \kappa^2 \frac{s_{34} + s_{14} }{16}.
\end{align}
In summary, the 4-scalar interaction term can be extracted as 
\begin{align}
	\mathcal{L}_{\varphi^4} =  \sqrt{-g} \frac{\kappa^2}{32} \varphi_\alpha^\dagger \varphi_\alpha D_\mu D^\mu (\varphi_\beta^\dagger \varphi_\beta).
\end{align}
The 6-scalar contact term is computed in the same way. For simplicity, the 3 pairs of scalars are taken to be of different flavors. Cases where two or three pairs of scalar are of the same flavor can be gained by simply exchanging the external particles. We match the amplitude computed from summing all Feynman diagrams to the double copy result. All the diagrams, except for the one from the 6-scalar interaction, are categorised into the six structures in fig. \ref{fig:image2}.
Subtracting every diagram that falls into either category in fig. \ref{fig:image2} from the double copy amplitude will give us the 6-scalar contact term,
\begin{align}
	\includegraphics[valign=c,scale=0.8]{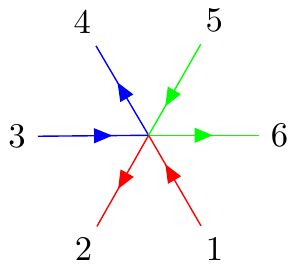} =  \frac{i \kappa^4}{256} \left( s_{12}+ s_{34} +s_{56} \right).
\end{align}
The corresponding interaction will be 
\begin{align}
	\mathcal{L}_{\varphi^6} = - \sqrt{-g} \frac{\kappa^4}{512} (\varphi_\alpha^\dagger \varphi_\alpha)^2 D_\mu D^\mu (\varphi_\beta^\dagger \varphi_\beta).
\end{align}
We now have calculated every term in \eqref{eq:L_pertubative}.

\section{Field redefinitions and a resummed action}

The established higher order terms in the action \eqref{eq:L_pertubative} do look 
somewhat non-standard with the double derivative $D^{\mu}D_{\mu}$ appearing. We would now
like to unify them with the kinetic and mass term in \eqref{eq:L_pertubative} by
performing a suitable field redefinition\footnote{We thank Alexander Ochirov for crucial discussions
on this point.}. To this end we consider the shift
\begin{equation}
\varphi_{\alpha} \to \varphi_{\alpha} + \frac{\kappa^{2}}{32} \varphi_{\alpha} 
\, (\varphi_{\beta}^{\dagger}\varphi_{\beta})
+ \frac{\kappa^{4}}{1024} \varphi_{\alpha} 
\, (\varphi_{\beta}^{\dagger}\varphi_{\beta})^{2}
\end{equation}
that transforms the matter part of the Lagrangian \eqref{eq:L_pertubative} into 
\begin{align}
\mathcal{L}_{\text{DCmatter}} &= 
\sqrt{-g} \bigg[ g^{\mu\nu} \partial_{\mu} \varphi_\alpha^\dagger \partial_\nu \varphi_\alpha - m_\alpha^2 e^{-\kappa \phi} \varphi_\alpha^\dagger \varphi_\alpha \bigg] \nonumber \\
	&\times \bigg [ 1 + \frac{\kappa^{2}}{16}  (\varphi_{\beta}^{\dagger}\varphi_{\beta})
	+ \frac{3\kappa^{4}}{1024}  (\varphi_{\beta}^{\dagger}\varphi_{\beta})^{2} +
	\cO(\kappa^{6})
	\bigg]\, .
\end{align}
By the S-matrix equivalence theorem this action after the field redefinitions
and the previous one \eqref{eq:L_pertubative} have identical scattering amplitudes.
\footnote{For this the asymptotic equivalence of the two fields $\varphi_{\alpha}$ and 
$\varphi_{\alpha}' =\varphi_{\alpha} + \delta \varphi_{\alpha}$ is required, i.e.~it suffices
if $\delta \varphi_{\alpha}$ is higher order in fields (and couplings).} This
form of the action suggests itself to an attractive resummation into the compact form
\begin{align}
\mathcal{L}_{\text{DC}} &= \mathcal{L}_{\text{adg}}+
\sqrt{-g}\frac{ \partial_{\mu} \varphi_\alpha^\dagger \partial^\mu \varphi_\alpha - m_\alpha^2 e^{-\kappa \phi} \varphi_\alpha^\dagger \varphi_\alpha}{
 \left ( 1 - \frac{\kappa^{2}}{32}  \varphi_{\beta}^{\dagger}\varphi_{\beta}	\right)^{2}}\, ,
\end{align}
with $\mathcal{L}_{\text{adg}}$ the two-form-dilaton-gravity theory of \eqn{eq:L_ADG}.
 Note
that in 4D upon dualizing the two-form to an axion $\chi$ the axio-dilaton system displays
a striking similarity to the massive flavored scalar Lagrangian above. In 4D the double
copy of scalar QCD takes the form
\begin{align}
\mathcal{L}_{\text{$($sQCD$)^{2}$}} &= -\frac{2\sqrt{-g}}{\kappa^{2}} R+ \sqrt{-g}\frac{\partial_{\mu}\bar Z
\partial^{\mu}Z}{(1- \frac{\kappa^{2}}{4}\bar Z Z)^{2}} \nn\\ &
+\sqrt{-g}\frac{ \partial_{\mu} \varphi_\alpha^\dagger \partial^\mu \varphi_\alpha - m_\alpha^2 e^{-\kappa \phi} \varphi_\alpha^\dagger \varphi_\alpha}{
 \left ( 1 - \frac{\kappa^{2}}{32}  \varphi_{\beta}^{\dagger}\varphi_{\beta}	\right)^{2}}\, .
\end{align}
Here the complex scalar field $Z$ is built from the dilaton and axion as
%
%
\begin{align}
	Z = \frac{2}{\kappa} \frac{{ \kappa }\chi+i(e^{-\kappa \phi}-1)}{{ \kappa } \chi+i(e^{-\kappa \phi}+1)}
\end{align}
and enjoys an $\text{SL}(2,\mathbf{R})$ symmetry (see e.g.~\cite{Schwarz:1992tn}). 
We note in closing that such a 
symmetry is also present in the scalar sector for the $N_{f}=1$  and \emph{massless} case.

\section{Conclusions}

In this letter we explicitly constructed the double copy of scalar QCD in arbitrary dimensions resulting in an extension
of the established two-form-dilaton gravity model (``$\mathcal{N}=0$ supergravity'') by interacting massive
flavored scalars displaying self interactions to arbitrary field orders. We found no coupling of the flavored scalars
to the two-form, i.e.~axion in 4d, as is commonly expected, yet not proven. Moreover, we saw that the 
self-interaction terms are always quadratic in derivatives. 
We constructed the emerging Lagrangian explicitly 
up to sixth order in fields. It would be important to develop a deeper understanding as to what symmetry
controls the form of these self-interactions at all orders. It should be an extension of the still elusive 
kinematic algebra.  We expect these results to be of value in the future for using the 
 double-copy
formalism in the two-body problem in classical General Relativity which should emerge from the infinite mass limit
of the theory we have constructed. Yet it appears to us that the established
novel contact terms will not alter the classical limit of the two massive scalar scatterings and cannot account for the discrepancies observed in \cite{Plefka:2019hmz}.

\acknowledgements
We wish to thank Alexander Ochirov for illuminating discussions leading to the resummed
form of the scalar action.
This project has received funding from the European Union's Horizon 2020 research and innovation program under the Marie Sklodowska-Curie grant agreement No. 764850,
and the German Research Foundation through grant PL457/3-1.

\appendix
\section{Feynman Rules}\label{appendix:FeynRules}
The covariant derivative and the field strength tensor in \eqref{eq:LsQCD} read
\begin{gather}
	D_\mu \varphi_{\alpha,i} = \partial_\mu \varphi_{\alpha,i} - i \frac{g}{\sqrt 2} A_\mu^a T^a_{ij} \varphi_{\alpha,j},\ \quad \\
	F^a_{\mu\nu} = \partial_{\mu} A_{\nu}^{a}-\partial_{\nu} A_{\mu}^{a} - i \frac{g}{\sqrt 2}f^{a b c} A_{\mu}^{b} A_{\nu}^{c}.
\end{gather}
The generators of the gauge group are normalized such that
\begin{gather}
	[T^a , T^b]_{ij} = f^{abc} T^c_{ij},\\
	\text{Tr}(T^a T^b)= \delta^{ab}.
\end{gather} 
The Lagrangian \eqref{eq:LsQCD} gives canonically normalized propagators and interaction vertices involving only gluons. The couplings of massive scalars with gluons read
\begin{align}
\includegraphics[valign=c, scale=0.7]{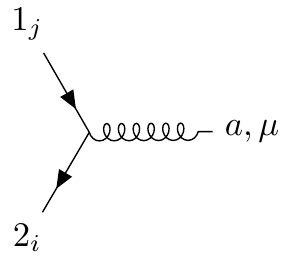}
&= i \frac{g}{\sqrt{2}} T^a_{ij} (p_1-p_2)^\mu, \\
\includegraphics[valign=c, scale=0.6]{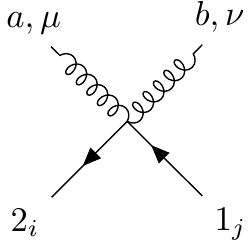}
&= i \frac{g^2}{2} (T^a_{ik} T^b_{kj}+T^b_{ik}T^a_{kj})\eta^{\mu\nu}.
\end{align}

As for the gravitational theory \eqref{eq:L_pertubative}, we adopt the de Donder gauge, where the propagators are
\begin{gather}
\includegraphics[scale=0.75]{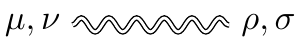} = \frac{i}{p^2} \left(\eta^{\mu(\rho} \eta^{\sigma)\nu} \!-\! \frac{1}{D-2} \eta^{\mu \nu} \eta^{\rho \sigma} \right),\!\!  \\
\includegraphics[scale=0.75]{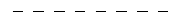} = \frac{i}{(D-2) p^2}.
\end{gather}
Since the kinetic term of dilaton is non-canonical, we dress each external dilaton by a factor ${1/ \sqrt{D-2}}$. The propagator of B-field is not needed in this paper. The Feynman rules of massive scalars coupled to graviton and dilaton are
respectively
\begin{gather}
	\includegraphics[valign=c,scale=0.7]{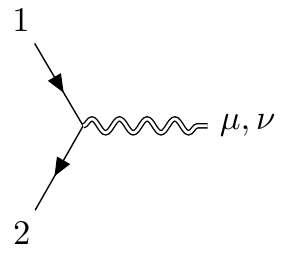} = -i \kappa \left( \frac{\eta^{\mu\nu}}{2} (p_1 \cdot p_2 + m^2) - p_1^{(\mu} p_2^{\nu)}  \right), \\
	\includegraphics[valign=c,scale=0.7]{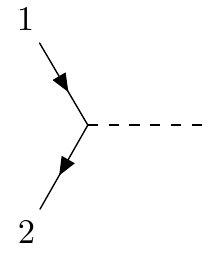} =  i \kappa m^2
\end{gather}
\begin{equation}
\begin{split}
    \includegraphics[valign=c,scale=0.7]{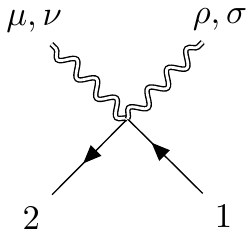} =&
    \begin{aligned}
         &\\&\frac{i \kappa ^2}{2}  \Big[ \eta^{\rho \sigma } p_1^{(\nu} p_2^{\mu)} + \eta^{\mu \nu} p_1^{(\sigma} p_2^{\rho)}  \\
         & - 2 p_1^{(\sigma} \eta^{\rho) (\mu } p_2^{\nu)} - 2 p_1^{(\mu} \eta^{\nu) (\sigma} p_2^{\rho)} 
    \end{aligned}\\
    + \Big( m^2 &+ p_1  \cdot p_2 \Big) \Big(\eta^{\mu(\rho} \eta^{\sigma) \nu} - \frac{1}{2} \eta^{\mu \nu } \eta^{\rho \sigma } \Big) \Big] 
\end{split}
\end{equation}
\begin{align}
	\includegraphics[valign=c,scale=0.8]{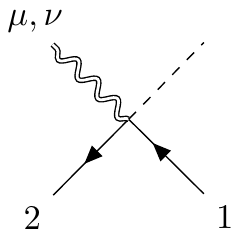} =  \frac{i \kappa^2 m^2 \eta_{\mu\nu}}{2}
\end{align}
\begin{align}
    \includegraphics[valign=c,scale=0.8]{./graphs/2massive2dilaton} =  -i \kappa^2 m^2
\end{align}
\begin{equation}
	\includegraphics[valign=c,scale=0.7]{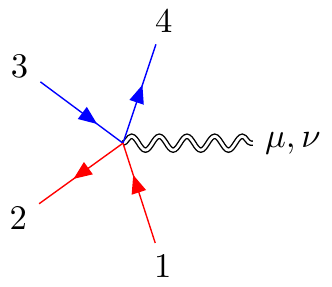} = 
	\begin{aligned}
		\frac{i \kappa ^3}{32} \Big[ (p_1+p_2) \cdot (p_3 + p_4) \eta^{\mu \nu}\\
		- 2 (p_1^\mu + p_2^\mu ) (p_3^\nu + p_4^\nu )  \Big]
	\end{aligned}
\end{equation}

The kinematic numerators in \eqref{eq:6scalars} are
\begin{align}
n_0 &= \frac{ig^4}{4} (p_1-p_2)_\mu (p_3-p_4)_\nu (p_5-p_6)_\rho\\ &\qquad\qquad V^{\mu\nu\rho}_{p_1+p_2, p_3+p_4, p_5+p_6}\\
n_1 &= -ig^4\Big[p_1\cdot (p_5-p_6)\, p_2\cdot(p_3-p_4) \nonumber\\
& \qquad\qquad+\frac{1}{4}(s_{134}-m_\alpha^2) (p_3 -p_4)\cdot (p_5-p_6) \Big] \!\! \\
n_2 &= -ig^4\Big[p_1\cdot (p_3-p_4)\, p_2\cdot(p_5-p_6) \nonumber\\
& \qquad\qquad+\frac{1}{4}(s_{156}-m_\alpha^2) (p_3 -p_4)\cdot (p_5-p_6) \Big] \!\!
\end{align}
where $$V^{\mu\nu\rho}_{pqk}=\eta^{\mu\nu}(p-q)^\rho+\text{cyclic}.$$ The other numerators are gained from cyclic rotations of the labels $(1,2)\rightarrow (3,4) \rightarrow (5,6)$ in $n_1$ and $n_2$.

\bibliographystyle{apsrev4-1}
\bibliography{inspire}
\end{document}